\documentclass[english,PRE,twocolumn]{revtex4-2}

\usepackage[T1]{fontenc}
\usepackage[latin9]{inputenc}
\usepackage{color}
\usepackage{amsbsy}
\usepackage{amstext}
\usepackage{amssymb}
\usepackage{graphicx}
\usepackage{esint}
\usepackage{babel}

\makeatletter
\usepackage{marginnote}

\makeatother

\begin{document}
\title{A geometric derivation of Noether's theorem}
\author{Bahram Houchmandzadeh.}
\address{CNRS, LIPHY, F-38000 Grenoble, France~\\
Univ. Grenoble Alpes, LIPHY, F-38000 Grenoble, France\\
bahram.houchmandzadeh@univ-grenoble-alpes.fr}

\begin{abstract}
Nother's theorem is a cornerstone of analytical mechanics, making
the link between symmetries and conserved quantities. In this article,
I propose a simple, geometric derivation of this theorem that circumvent
the usual difficulties that a student of this field usually encounters.
The derivation is based on the integration the differential form $dS=\mathbf{p}d\mathbf{q}-Hdt$,
where $S$ is the action function, $\mathbf{p}$ the momentum and
$H$ the Hamiltonian, over a closed path. 
\end{abstract}
\maketitle

\section{Introduction.}

Noether's theorem \cite{noether1971} is one of the most celebrated
theorem in physics, linking symmetries to conserved quantities. 

Early in their education, physics undergraduates learn that the well-known
conservation laws---momentum, energy, and angular momentum---are
connected to symmetries in nature, such as translation in space and
time, as well as rotation. This important relationship applies across
all areas of physics and was established by the mathematician Emmy
Noether in 1918. However, despite this beauty, they discover that
most classical textbooks on mechanics don't mention this theorem\cite{lanczosVariationalPrinciplesMechanics2012,feynman2015},
or devote a very small place to it\cite{landauMechanicsVolume1976,arnoldMathematicalMethodsClassical1997},
or give a limited version of it \cite{morin2009} or postpone it
to the very last section\cite{goldsteinClassicalMechanics2013} (for
an exception, see \cite{calkinLagrangianHamiltonianMechanics1996}
). 

There are excellent books\cite{neuenschwander2017} dedicated to
Noether's theorem. Moreover, there are no shortage of articles in
physics journal \cite{levy-leblond1971,desloge1977,hanc2004,gorni2014}
and public resources such as Wikipedia, to which the interested student
can turn to. There are however several pitfalls for students before
they can grasp the depth of Noether's theorem. The symmetry of Noether's
theorem is not the symmetry of the potential energy, not even the
symmetry of the Lagrangian ${\cal L}$ (which is harder to grasp,
as the latter evaluates over \emph{trajectories}, not point of space),
but the invariance of ${\cal L}dt$ under a transformation. The classical
examples of translation and rotation symmetry are slightly misleading,
because the student could get the impression that these are simply
those of the of potential energy.

The classical examples of applications of Noether's theorem are also
slightly underwhelming, as the usual conservation laws can be obtained
simply by the method of cyclic or ignorable variables, where the shape
of the potential is a guide for the choice of coordinates. Finally,
when all the above pitfalls are overcome, students have to digest
the heavy developments of the Rund-Trautman identity, making sense
of derivation in respect to time versus the transformation parameter,
apply that to an optimum trajectory, and discover that ``miraculously'',
most of the terms cancel out and lead to a simple expression where
the momentum $\mathbf{p}$ and the Hamiltonian $H$ appear in a symmetric
form reminiscent of relativity. For completeness, the Rund-Trautman
derivation is detailed in subsection\textcolor{red}{{} }\ref{subsec:Lagrangian-approach.}

These are in my opinion some of the reasons why most physics students
are unaware of the meaning of this beautiful theorem. The Noether's
theorem however is so elegant, has such a deep meaning and makes connections
between so many fields of physics that I believe it should be part
of the background of any advanced undergraduate or graduate student.
The theorem is also a first glimpse for students at Lie's theory of
continuous transformations, which was introduced in the 1880 precisely
to study systematically the classification of differential equations
according to their symmetries\cite{waerden1985}.

To overcome the difficulties enumerated above, I propose an alternative
and simple derivation of the Noether's theorem, highlighting its \emph{geometrical}
meaning. The derivation is based on the fundamental theorem of calculus
that states that, for an analytical function $S$ 
\begin{equation}
\oint dS=0\label{eq:analysis:fundmntl}
\end{equation}
where the integration is taken on a curve ${\cal C}$ inside the domain
where $S$ is analytical (figure \ref{fig:int:dS}). The function
$S\text{(\textbf{q}},t)$ here is the action \emph{function} that
is the solution of the Hamilton-Jacobi (HJ) equation, \emph{i.e.}
the functional $S$ computed over actual trajectories, where 
\begin{equation}
dS=\mathbf{p}d\mathbf{q}-Hdt\label{eq:dS}
\end{equation}
The curve ${\cal C}$ is formed of four branches, two of them being
actual trajectories, where one is the transformed of the other under
the transformation $(\phi_{\mathbf{q}},\phi_{t})$. If we suppose
that the integral over these two branches are equal (to first order
in a small parameter $\epsilon$), we must conclude that the integral
over the two other branches must also be equal, leading naturally
to Noether's theorem that 
\begin{equation}
I=\mathbf{p}\phi_{\mathbf{q}}-H\phi_{t}\label{eq:noether:charge}
\end{equation}
is a conserved quantity along the trajectory. There is nothing miraculous
in the simplifications of Rund-Trautman identity, these are a simple
consequence of the fundamental theorem of calculus (\ref{eq:analysis:fundmntl}).
This derivation is given in subsection \ref{subsec:H-J-Approach}.
To be complete, the derivation using the Rund-Trautman identity is
given in subsection \ref{subsec:Lagrangian-approach.}. Before getting
there however, I recall in section \ref{sec:Point-transformation}
the essential concepts of transformation groups and infinitesimal
generators that are needed for this derivation. These concepts are
the basis for the Lie theory of continuous transformations but can
be presented simply without the full development of Lie groups. Section
\ref{sec:Formulations-of-mechanics} is devoted to recalling the essential
ingredients of analytical mechanics and its different formulations,
mainly the Lagrangian and HJ formulation. Section \ref{sec:Conclusions}
is devoted to final concluding remarks. An appendix contains miscellaneous
details that are recalled here for clarity. 

\section{Point transformation and invariance\protect\label{sec:Point-transformation}}

\subsection{Definitions.}

A continuous, one parameter transformation $T_{s}$ ($s\in\mathbb{R}$)
transforms each point of the space $\boldsymbol{\mathbf{r}}$ into
a new one 
\begin{equation}
\boldsymbol{\mathbf{r}}'=T_{s}(\boldsymbol{\mathbf{r}})\label{eq:transformation}
\end{equation}
We restrict the discussion here to the case where transformations
$T_{s}$ form a \emph{group} \cite{zee2016}. Prime examples of such
transformations are translations and rotations. 

In order to characterize a transformation $T_{s}$, we only need to
know its \emph{infinitesimal generator} , \emph{i.e.} how points are
transformed for infinitesimal values of the parameter $s$. For example,
in a two dimensional space referred to by the euclidean coordinates
$(x,y)$, an infinitesimal rotation $\epsilon$ around the origin
is characterized by 
\begin{eqnarray}
x' & = & x-\epsilon y\label{eq:rotx}\\
y' & = & y+\epsilon x\label{eq:roty}
\end{eqnarray}
In vectorial notation, we write the infinitesimal transformation as
\begin{equation}
\boldsymbol{\mathbf{r}}'=\boldsymbol{\mathbf{r}}+\epsilon\boldsymbol{\phi}(\boldsymbol{\mathbf{r}})\label{eq:infinitesimal:generator}
\end{equation}
where the vector field $\boldsymbol{\phi}(\mathbf{r})$ is called
the infinitesimal generator of the the transformation $T_{s}$ (figure
\ref{fig:vector-field}). The vector field $\boldsymbol{\phi}(\mathbf{r})$
entirely characterizes the transformation $T_{s}$ : if we want to
know the transformation for a finite value of $s$, we need to apply
the transformation $T_{\epsilon}$ repeatedly, which is equivalent
to solving the differential equation 
\begin{equation}
\frac{d\boldsymbol{\mathbf{r}}}{ds}=\boldsymbol{\phi}(\boldsymbol{\mathbf{r}})\label{eq:difeq}
\end{equation}
Solving the above equation for example for rotations (equations (\ref{eq:rotx},\ref{eq:roty})
), we find the usual expression for 2d rotations : 
\begin{figure}
\begin{centering}
\includegraphics[width=0.5\columnwidth]{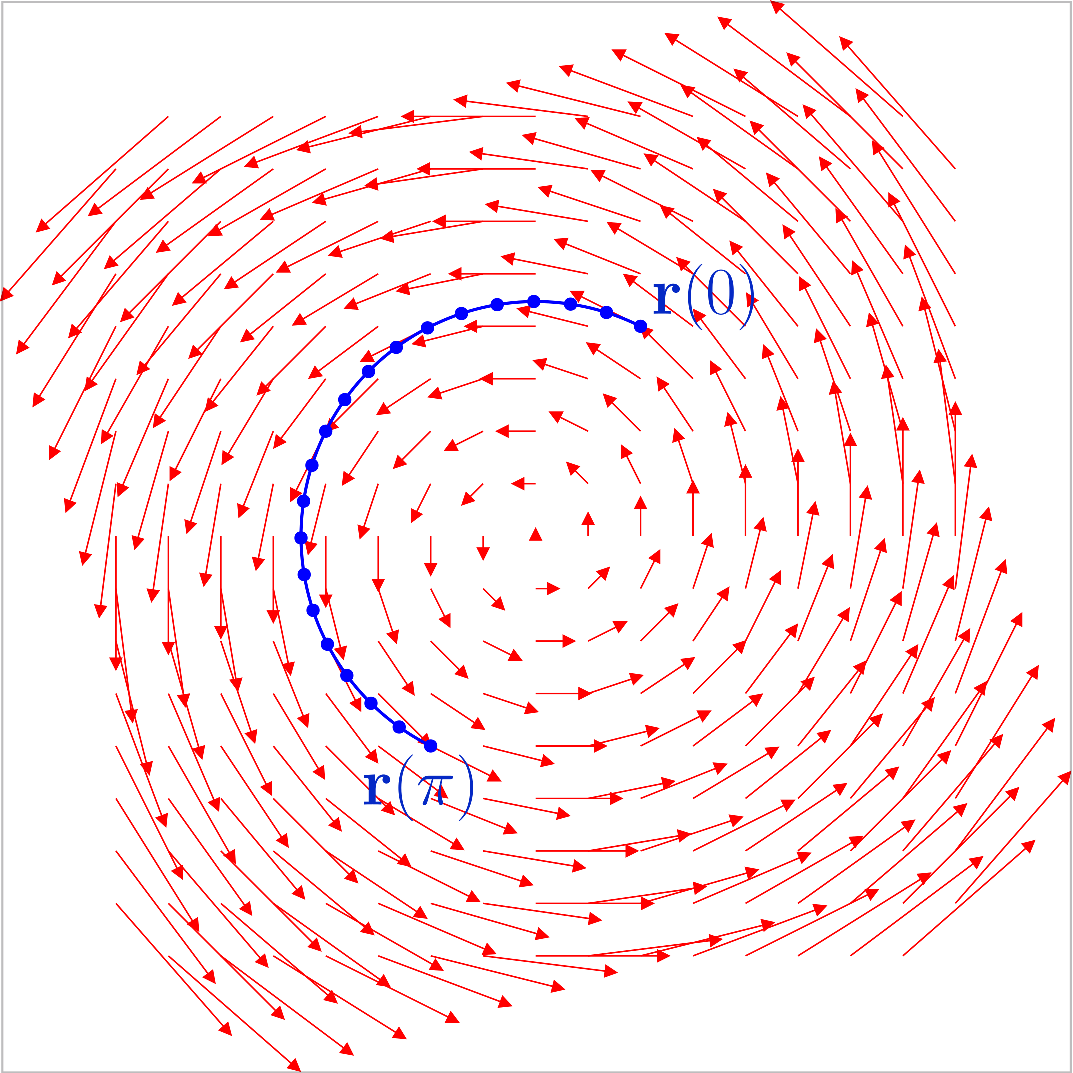}
\par\end{centering}
\caption{A vector field (red arrow) that transform (infinitesimally) each point
into another $\boldsymbol{\mathbf{r}}'=\boldsymbol{\mathbf{r}}+\epsilon\boldsymbol{\phi}(\boldsymbol{\mathbf{r}})$.
In this example, the field corresponds to a rotation. To rotate by
a finite quantity $\theta$, the infinitesimal transformation is applied
$M$ times, where $M\epsilon=\theta$. This is equivalent to solving
the differential equation (\ref{eq:difeq}), or finding the curve
that is tangent to the vector field at each point (blue curve).\protect\label{fig:vector-field} }

\end{figure}
\begin{eqnarray}
x(s) & = & x(0)\cos(s)-y(0)\sin(s)\label{eq:rot:s:1}\\
y(s) & = & x(0)\sin(s)+y(0)\cos(s)\label{eq:rot:s:2}
\end{eqnarray}

\subsection{Symmetry and invariance.}

Consider a function $f(\mathbf{r})$ defined over the points of space
$\mathbf{r}=(r^{1},\ldots r^{n})$. The function is said to be \emph{invariant}
under a transformation $T_{s}$ if for every point of space, 
\begin{equation}
f(T_{s}\mathbf{r})=f(\mathbf{r})\label{eq:invariance:global}
\end{equation}
For the relation (\ref{eq:invariance:global}) to be valid, we only
need it to be valid for an infinitesimal transformation : 
\begin{equation}
f\left(\mathbf{r}+\epsilon\boldsymbol{\phi}(\boldsymbol{\mathbf{r}})\right)=f(\mathbf{r})\label{eq:invariance:local-1}
\end{equation}
Developing to the first order in $\epsilon$, we then must have 
\begin{equation}
\sum\frac{\partial f}{\partial r^{i}}\phi^{i}(\mathbf{r})=\left(\frac{\partial f}{\partial r^{1}},\ldots,\frac{\partial f}{\partial r^{n}}\right)\left(\begin{array}{c}
\phi^{1}\\
\vdots\\
\phi^{n}
\end{array}\right)=0\label{eq:invariance:local}
\end{equation}
The application of the linear form $(\partial_{1}f,\ldots,\partial_{n}f)$
to the vector $\boldsymbol{\phi}$ must be null. The relation (\ref{eq:invariance:local})
is the condition for a function to be invariant under a transformation
$\phi$. 

For a given transformation $\phi$, solving the partial differential
equation (\ref{eq:invariance:local}) gives the family of function
that are invariant under this transformation. For example, consider
a function $f(x,y)$ that is invariant under rotations. The infinitesimal
generator of rotations is the vector field $\mathbf{\phi}=(-y,x)$,
therefore we must have 
\begin{equation}
-\frac{\partial f}{\partial x}y+\frac{\partial f}{\partial y}x=0\label{eq:invariance:local:rotation}
\end{equation}
It is straightforward to check that any function of the form $f(x,y)=g(x^{2}+y^{2})$
is a solution of equation (\ref{eq:invariance:local:rotation}). 

In classical mechanics, one of the coordinates, usually called time
$t$, is distinguished from all the others called space $\mathbf{q}$.
The vector field of a transformation is broken into its space and
time components $\boldsymbol{\phi}(\mathbf{r})=\left(\boldsymbol{\phi}_{\mathbf{q}};\phi_{t}\right)^{T}$
; this complicated notation better corresponds to our everyday experience,
where we measure space by a stick and time by a clock. In the following
sections, we will use this convention of classical mechanics. 

\section{Formulations of mechanics\protect\label{sec:Formulations-of-mechanics}}

There are three different and equivalent formulations of analytical
mechanics. We recall below the Lagrangian (see for example \cite{lanczosVariationalPrinciplesMechanics2012,goldsteinClassicalMechanics2013})
and the Hamilton-Jacobi formulations (see for example \cite{arnoldMathematicalMethodsClassical1997},
p248 or \cite{houchmandzadehHamiltonJacobiEquation2020}). The Hamilton
description, obtained through a Legendre transform of the Lagrangian,
is not necessary for the present discussion of Noether's theorem and
is omitted. 

\subsection{Lagrangian formulations}

The first approach to analytical mechanics is the Lagrangian one.
In this formulation, a system is specified by a Lagrangian ${\cal L}(\mathbf{q},\dot{\mathbf{q}},t)$,
where $\mathbf{q}(t)=\left(q^{1}(t),\ldots q^{n}(t)\right)^{T}$ is
the position of the system at time $t$ in a given system of coordinates
and $\dot{\mathbf{q}}=d\mathbf{q}/dt$ is its generalized velocity.
To find the trajectory $\mathbf{q}(t)$ of the system one defines
the momentum 
\begin{equation}
\mathbf{p}=\frac{\partial{\cal L}}{\partial\dot{\mathbf{q}}}\label{eq:momentum:def}
\end{equation}
and then uses the Euler-Lagrange equation 
\begin{equation}
\frac{d}{dt}\mathbf{p}=\frac{\partial{\cal L}}{\partial\mathbf{q}}\label{eq:EL:def}
\end{equation}
to find the trajectories. In most problems of classical mechanics,
equation (\ref{eq:EL:def}) is a system of second order differential
equations in $q^{i}(t)$. 

It is important to precise the notations used here. Equation (\ref{eq:momentum:def})
is a short hand notation for 
\begin{equation}
p_{i}=\frac{\partial{\cal L}}{\partial\dot{q}^{i}}\label{eq:moment:indexes}
\end{equation}
and $\mathbf{p}=(p_{1},\ldots,p_{n})$. Note that $\mathbf{p}$ is
not a vector, but a \emph{linear form}. In Linear algebra, vectors
components are often grouped into a column vector, while linear forms
are represented by a row vector. Here we make the distinction by using
upper indexes for vector components (contravariant vector) and lower
indexes for linear forms (covariant vectors). Applications of a linear
form to a vector produces a scalar. For example, 
\begin{equation}
\mathbf{p}\dot{\mathbf{q}}=\sum p_{i}\dot{q}^{i}\label{eq:moment:apply}
\end{equation}
An important \emph{scalar }quantity that is deduced from the Lagrangian
is the \emph{Hamiltonian}
\begin{equation}
H=\mathbf{p}\dot{\mathbf{q}}-{\cal L}\label{eq:Hamiltonian:def}
\end{equation}
From the above relation, it is straightforward to deduce that (see
subsection \ref{subsec:Hamiltonian-variation})
\begin{equation}
\frac{dH}{dt}=-\frac{\partial{\cal L}}{\partial t}\label{eq:dH:dt}
\end{equation}
if the Lagrangian does not contain time explicitly, then the Hamiltonian
is conserved along a trajectory. 

As an example, consider a two dimensional harmonic oscillator, where
the Lagrangian is given by 
\begin{equation}
{\cal L}=\frac{m}{2}\left(\dot{x}^{2}+\dot{y}^{2}\right)-\frac{k}{2}\left(x^{2}+y^{2}\right)\label{eq:harmonic:osc:lagrangian}
\end{equation}
The momentum is $\mathbf{p}=m(\dot{x},\dot{y})$ and the Hamiltonian
is therefore
\begin{equation}
H=\frac{m}{2}\left(\dot{x}^{2}+\dot{y}^{2}\right)+\frac{k}{2}\left(x^{2}+y^{2}\right)\label{eq:harmonic:osc:hamiltonian}
\end{equation}

As we will see in the next section, the generalized $n+1$ linear
form 
\begin{equation}
(\mathbf{p},-H)\label{eq:generlized:momentum}
\end{equation}
plays the central role in the Noether's theorem : If the vector field
$\left(\mathbf{\phi_{\mathbf{q}}};\phi_{t}\right)^{T}$ is the infinitesimal
generator of a symmetry of the system, the conserved quantity along
a trajectory is the scalar 
\begin{equation}
(\mathbf{p},-H)\left(\begin{array}{c}
\mathbf{\phi_{\mathbf{q}}}\\
\phi_{t}
\end{array}\right)=\mathbf{p}\phi_{\mathbf{q}}-H\phi_{t}\label{eq:noether:charge-1}
\end{equation}
Note that, as mentioned in the previous section, the term ``generalized''
is an exaggeration. Classically, one coordinate, called time $t$,
is distinguished from the others, called space $\mathbf{q}$ and therefore
various objects such as vectors and linear forms are broken into their
time and space components, hiding the symmetry of the notations. The
linear form $(\mathbf{p},-H)$ is widely used in special relativity
: its transpose $(\mathbf{p},H)^{T}$ for a particle is called the
four-vector momentum ; The sign (-) is the signature of our space-time. 

Note also that the Lagrangian is not unique and is defined up a to
a total derivative. If ${\cal L}(\mathbf{q},\dot{\mathbf{q}},t)$
describes a system and gives its trajectories, the Lagrangian 
\begin{equation}
{\cal L}^{\dagger}(\mathbf{q},\dot{\mathbf{q}},t)={\cal L}(\mathbf{q},\dot{\mathbf{q}},t)+\frac{d}{dt}F(\mathbf{q},t)\label{eq:lagrangian:total:derivative}
\end{equation}
describes the same system and gives the same trajectories (see 
\ref{subsec:Non-unicity}). This indeterminacy in the Lagrangian introduces
also an additional term in the most general statement of Noether's
theorem, which we will consider in  \ref{subsec:Non-unicity}. 

\subsection{Hamilton-Jacobi formulation.\protect\label{subsec:Hamilton-Jacobi-formulation.}}

The Euler-Lagrange equation (\ref{eq:EL:def}) is a consequence of
an ``extremum'' concept. Given a \emph{functional} $S$ of trajectories,
called the ``\emph{action}''
\begin{equation}
S[\mathbf{q}(t)]=\int_{A}^{B}{\cal L}(\mathbf{q},\dot{\mathbf{q}},t)dt\label{eq:action1:def}
\end{equation}
where $A=(\mathbf{q}_{0},t_{0})$ and $B=(\mathbf{q}_{1},t_{1})$
are two end points of the curves, the trajectory followed by a system
is one that corresponds to the extremum of the action. Such a trajectory
is shown to obey the Euler-Lagrange equation (\ref{eq:EL:def}). 

On the other hand, if we \emph{know} the real trajectories, $S=S(\mathbf{q}_{1},t_{1})$
becomes a simple function of the points in space, given an initial
condition. The function $S$ is computed by evaluating the integral
(\ref{eq:action1:def}) over the known trajectories, themselves computed
from the Euler-Lagrange equation (\ref{eq:EL:def}). We can envisage
$S(\mathbf{q}_{1},t_{1})$ as a \emph{wave front} at time $t_{1}$
in analogy with geometrical optics. 

As in the case of geometrical optics (which is a particular case of
analytical mechanics), there is a duality between trajectories and
wave fronts\cite{houchmandzadehHamiltonJacobiEquation2020}. For
optics, the duality was first developed by Huygens \cite{shapiro1989}
in 1690 and it was fully integrated into analytical mechanics by Hamilton
and Jacobi\cite{dugasHistoryMechanics2012}. 

By duality, we mean that if we \emph{know} the the action function
$S(\mathbf{q},t)$ at all points of the space, we can \emph{deduce}
the trajectories (figure \ref{fig:Computing-the-trajectories}). The
action \emph{function }is related to the generalized momentum (\ref{eq:generlized:momentum})
through the relations (see  \ref{subsec:Hamilton-Jacobi-equation})
: 
\begin{figure}
\begin{centering}
\includegraphics[width=0.7\columnwidth]{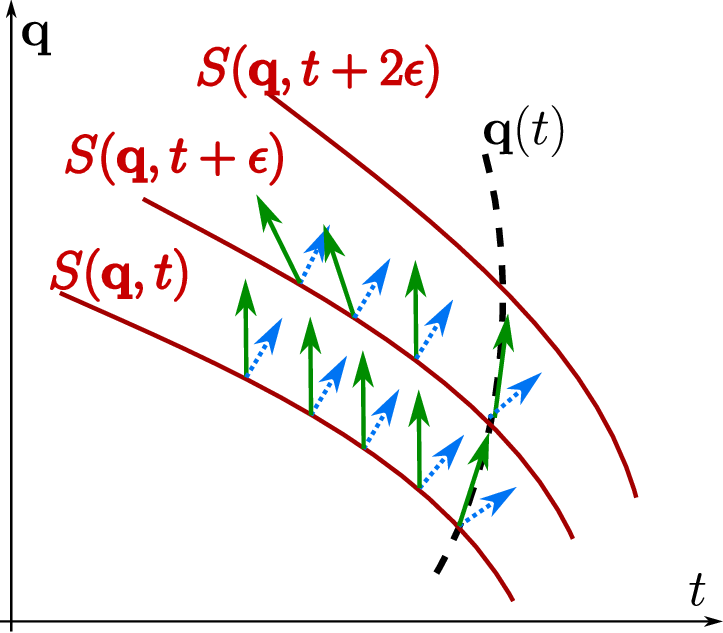}
\par\end{centering}
\caption{Computing the trajectories from the action $S(\mathbf{q},t)$ (here
schematized by a 2d red contour plot). At each point, the momentum
$(\mathbf{p},-H)=(\partial S/\partial\mathbf{q},\partial S/\partial t)$
(in blue) can be computed ; knowing $\mathbf{p}$, we can compute,
from relation (\ref{eq:momentum:def}) the tangent to the trajectory
$(\dot{\mathbf{q}},1)^{T}$ (in green). The trajectory (in black)
is found by following a given line of tangents. \protect\label{fig:Computing-the-trajectories}}
\end{figure}
\begin{eqnarray}
\frac{\partial S}{\partial\mathbf{q}} & = & \mathbf{p}\label{eq:action:momentum}\\
\frac{\partial S}{\partial t} & = & -H\label{eq:action:hamiltonian}
\end{eqnarray}
Therefore, at each point of space, $(\mathbf{p},-H)$ is known. Knowing
$\mathbf{p}$, we can find $\dot{\mathbf{q}}$ by solving the equation
(\ref{eq:momentum:def}), which is usually an algebraic expression
; knowing $\dot{\mathbf{q}}(\mathbf{q},t)$, we solve the first degree
ODE to find $\mathbf{q}(t)$ (figure \ref{fig:Computing-the-trajectories}). 

The action $S(\mathbf{q},t)$ itself is computed from a first order
nonlinear partial differential equation equation called the Hamilton-Jacobi
equation (see  \ref{subsec:Hamilton-Jacobi-equation}). The
Hamilton-Jacobi approach, in the words of Arnold\cite{arnoldMathematicalMethodsClassical1997},
is ``{[}...{]} the most powerful method known for the exact integration
{[}of Hamilton equations{]}''. The aim of this article is not to
study the H-J approach (see for more details for example \cite{houchmandzadehHamiltonJacobiEquation2020}).
For the discussion of Noether's theorem, we only need to know that
the function $S(\mathbf{q},t)$ exists, is computable and analytical.
Moreover, its differential, according to relations (\ref{eq:action:momentum},\ref{eq:action:hamiltonian})
is 
\begin{equation}
dS=\mathbf{p}d\mathbf{q}-Hdt\label{eq:action:differential}
\end{equation}
which is the linear form (\ref{eq:generlized:momentum}) we briefly
discussed in the previous subsection. 

\section{Noether's theorem}

There are various formulation of the Noether's theorem. The simplest
one, in our opinion, is the following: 

If the action 
\begin{equation}
S=\int_{A}^{B}{\cal L}(\mathbf{q},\dot{\mathbf{q}},t)dt\label{eq:action:def}
\end{equation}
remains invariant under an infinitesimal transformation 
\begin{equation}
\mathbf{\Phi=}(\mathbf{\phi}_{\mathbf{q}},\mathbf{\phi}_{t})\label{eq:transformation:symmetry}
\end{equation}
then the quantity 
\begin{equation}
I=(\mathbf{p},-H)\left(\begin{array}{c}
\mathbf{\phi_{\mathbf{q}}}\\
\phi_{t}
\end{array}\right)=\mathbf{p}\phi_{\mathbf{q}}-H\phi_{t}\label{eq:noether:charge:2}
\end{equation}
is conserved along the trajectory taken by the system. 

The invariance means that under the transformation 
\begin{eqnarray}
\mathbf{q}' & = & \mathbf{q}+\epsilon\mathbf{\phi_{\mathbf{q}}}\label{eq:phi:q}\\
t' & = & t+\epsilon\phi_{t}\label{eq:phi:t}
\end{eqnarray}
We must have $S'=S$ to the first order in $\epsilon$, or in other
words, 
\begin{equation}
{\cal L}(\mathbf{q}',\dot{\mathbf{q}'},t')dt'={\cal L}(\mathbf{q},\dot{\mathbf{q}},t)dt\label{eq:noether:condition}
\end{equation}
to the first order in $\epsilon$. This obviously implies that if
$\mathbf{q}(t)$ is an extremum of the functional $S$, then $\mathbf{q}'(t')$
is an extremum of $S'$ : the transformation (\ref{eq:transformation:symmetry})
transforms extremum trajectories into extremum trajectories. 

Let us precise the contour of Noether's theorem. This theorem allows
us, if we \emph{know} of a symmetry, to deduce the conserved quantities.
It does not allow us to \emph{find} the symmetries in a efficient
way ; for this purpose, there are other methods such as canonical
transformation or the HJE equation. Moreover, it states that to a
given symmetry a conserved quantity is associated. The converse however
is not generally true, all conserved quantities do not derive from
a symmetry, and not all transformations that transform a trajectory
into a trajectory conserve the Lagrangian\cite[subsection 4.20]{arnoldMathematicalMethodsClassical1997}.

Finally, the most general form of Noether's condition allows for the
Lagrangians difference to be equal to a total time derivative to the
first order in $\epsilon$ : 
\begin{equation}
{\cal L}(\mathbf{q}',\dot{\mathbf{q}'},t')\frac{dt'}{dt}={\cal L}(\mathbf{q},\dot{\mathbf{q}},t)+\epsilon\frac{d}{dt}F(\mathbf{q},t)\label{eq:noether:general:condition}
\end{equation}
in which case, the conserved quantity along a trajectory is 
\begin{equation}
\mathbf{p}\phi_{\mathbf{q}}-H\phi_{t}-F(\mathbf{q},t)\label{eq:noether:charge:general}
\end{equation}
This generalization is considered in  \ref{subsec:Non-unicity}. 

\subsection{H-J Approach\protect\label{subsec:H-J-Approach}}

\begin{figure}
\begin{centering}
\includegraphics[width=0.7\columnwidth]{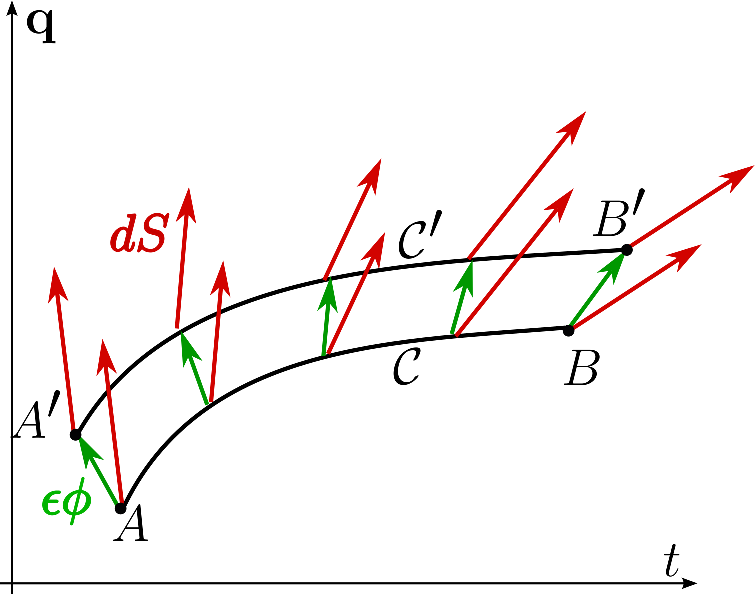}
\par\end{centering}
\caption{The integration path of $dS$ along a closed circuit ${\cal P}$.
In red, the linear form $dS=\mathbf{p}d\mathbf{q}-Hdt$. The green
vectors refer to the transformation $\Phi$.\protect\label{fig:int:dS}}

\end{figure}

Consider the action function $S(\mathbf{q},t)$ as defined in subsection
\ref{subsec:H-J-Approach}, \emph{i.e. }as the integral of the Lagrangian
from an initial point $(\mathbf{q}_{0},t_{0})$ to the point $(\mathbf{q},t)$
over optimal paths. The hypotheses of the Noether's theorem is equivalent
to stating that $S$ is invariant (up to an additive constant) under
the transformation field $\Phi=(\phi_{\mathbf{q}},\phi_{t})$. Now,
consider a given action function $S(\mathbf{q},t)$ and two close
trajectories ${\cal C}$ and ${\cal C}'$ that are associated to $S$,
by the procedure described in subsection \ref{subsec:Hamilton-Jacobi-formulation.},
and where ${\cal C}'$ is a transform of ${\cal C}$ by the field
$\Phi=(\phi_{\mathbf{q}},\phi_{t})$ (figure \ref{fig:int:dS}). Let
us define a closed path ${\cal P}=AA'B'BA$ where the branches $AB={\cal C}$
and $A'B'={\cal C}'$. 

By the virtue of the fundamental theorem of Calculus\cite{edwardsAdvancedCalculusDifferential1993}
(see also  \ref{subsec:Fundamental-Theorem}), we know that
for a function $S$ analytic in a domain ${\cal D}$, we have 
\begin{equation}
\oint dS=0\label{eq:integral:dS}
\end{equation}
where the integration path ${\cal P}$ is inside the domain ${\cal D}$.
By the hypothesis of Noether's theorem, we have 

\begin{equation}
\int_{{\cal C}}dS=\int_{{\cal C}'}dS\label{eq:int:along:branch}
\end{equation}
therefore, by virtue of theorem (\ref{eq:integral:dS}), we also must
have
\begin{equation}
\int_{A}^{A'}dS=\int_{B}^{B'}dS\label{eq:int:along:transform}
\end{equation}
Along the $AA'$ branch, $d\mathbf{q}=\epsilon\phi_{\mathbf{q}}$
and $dt=\epsilon\phi_{t}$. The integral along the $AA'$ branch,
to the first order in $\epsilon$, is then simply 
\begin{equation}
\int_{A}^{A'}dS=\epsilon\left(\mathbf{p}\phi_{\mathbf{q}}-H\phi_{t}\right)+o(\epsilon)\label{eq:int:along:transform:1order}
\end{equation}
Therefore, the quantity 
\begin{equation}
I=\mathbf{p}\phi_{\mathbf{q}}-H\phi_{t}\label{eq:noether:charge:again}
\end{equation}
is conserved along the trajectory. 

\subsection{Lagrangian approach.\protect\label{subsec:Lagrangian-approach.}}

The Noether's theorem in all sources I am aware of is demonstrated
using the Lagrangian approach. This demonstration is equivalent to
demonstrating that $dS$ is a closed form. We recall this demonstration
here.

\begin{figure}
\begin{centering}
\includegraphics[width=0.7\columnwidth]{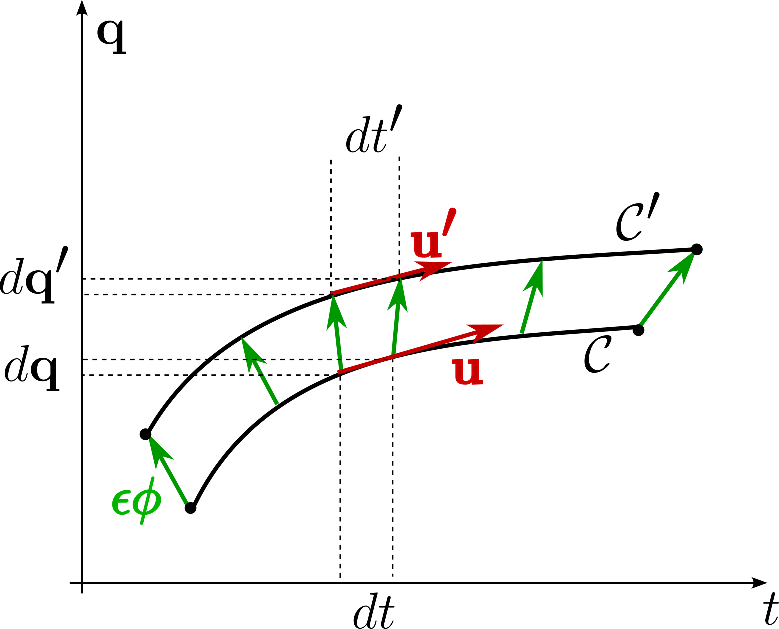}
\par\end{centering}
\caption{Scheme of transformation of a curve ${\cal C}$ into the curve ${\cal C}'$
by the vector field $\Phi$. the vector $\mathbf{u}=(\dot{\mathbf{q}},1)$
in red is the tangent vector to the curve ${\cal C}$. \protect\label{fig:Scheme-of-transformation} }

\end{figure}

Consider a curve ${\cal C}$ parameterized by the coordinate $t$
and written as $\mathbf{q}(t)$ (figure \ref{fig:Scheme-of-transformation}).
Consider a field of transformation $\mathbf{\Phi=}\left(\mathbf{\phi}_{\mathbf{q}}(\mathbf{q},t),\mathbf{\phi}_{t}(\mathbf{q},t)\right)$:
\begin{eqnarray}
\mathbf{q}' & = & \mathbf{q}+\epsilon\mathbf{\phi_{\mathbf{q}}}\label{eq:q'}\\
t' & = & t+\epsilon\phi_{t}\label{eq:t'}
\end{eqnarray}
Along the curve ${\cal C}$, $\Phi$ itself can be parameterized by
$t$
\begin{equation}
\Phi(t)=\Phi(\mathbf{q}(t),t)\label{eq:phi:of:t}
\end{equation}
and therefore, along the curve, the derivative $d\Phi/dt$ is well
defined. Consider now two close points along the curve ${\cal C}$
with parameter $t_{1}$ and $t_{2}=t_{1}+dt$. The image of these
points on ${\cal C}'$ are given by $t'_{1}=t_{1}+\epsilon\phi_{t}(t_{1})$
and 
\begin{equation}
t'_{2}=t_{2}+\epsilon\phi_{t}(t_{1}+dt)=t_{2}+\frac{d\phi_{t}}{dt}\epsilon dt\label{eq:t'2}
\end{equation}
Therefore 
\begin{equation}
dt'=t'_{2}-t'_{1}=\left(1+\epsilon\frac{d\phi_{t}}{dt}\right)dt\label{eq:dt'}
\end{equation}
By the same method, we obtain 
\begin{equation}
d\mathbf{q}'=d\mathbf{q}+\epsilon\frac{d\phi_{\mathbf{q}}}{dt}dt\label{eq:dq'}
\end{equation}
Therefore, the quantity 
\begin{equation}
\dot{\mathbf{q}}(t)=\frac{\mathbf{q}(t+dt)-\mathbf{q}(t)}{dt}\label{eq:dot:q:prime}
\end{equation}
transforms as 
\begin{equation}
\dot{\mathbf{q}}'(t)=\frac{d\mathbf{q}'}{dt'}=\dot{\mathbf{q}}(t)+\epsilon\left(\frac{d\phi_{\mathbf{q}}}{dt}-\dot{\mathbf{q}}\frac{d\phi_{t}}{dt}\right)\label{eq:qdot:prime}
\end{equation}
Recall that the hypothesis of the Nother's theorem is that 
\begin{equation}
{\cal L}(\mathbf{q}',\dot{\mathbf{q}'},t')dt'={\cal L}(\mathbf{q},\dot{\mathbf{q}},t)dt\label{eq:noether:hypothesis}
\end{equation}
The relations (\ref{eq:q'},\ref{eq:t'},\ref{eq:qdot:prime}) allow
us to expand relation (\ref{eq:noether:hypothesis}) to the first
order in $\epsilon$ and write
\begin{eqnarray}
\left[{\cal L}+\epsilon\left(\frac{\partial{\cal L}}{\partial\mathbf{q}}\mathbf{\phi_{\mathbf{q}}}+\frac{\partial{\cal L}}{\partial\dot{\mathbf{q}}}\left(\frac{d\phi_{\mathbf{q}}}{dt}-\dot{\mathbf{q}}\frac{d\phi_{t}}{dt}\right)+\frac{\partial{\cal L}}{\partial t}\phi_{t}\right)\right]\nonumber \\
\times\left(1+\epsilon\frac{d\phi_{t}}{dt}\right)={\cal L} &  & \,\,\,\,\,\,\,\,\,\,\,\,\label{eq:lagrangian:development}
\end{eqnarray}

\begin{figure}
\begin{centering}
\includegraphics[width=0.75\columnwidth]{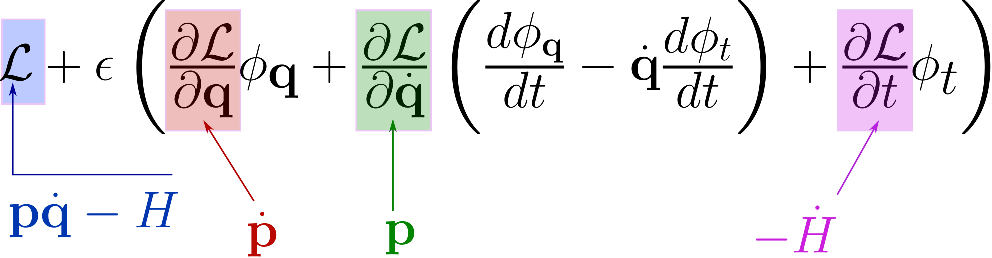}
\par\end{centering}
\caption{Scheme of replacements in expression (\ref{eq:lagrangian:development}).
\protect\label{fig:Scheme-of-replacements}}
\end{figure}

Now, we suppose that the curve ${\cal C}$ is a trajectory of the
system and obeys the Euler-Lagrange equation. Therefore, in the above
relation, we can replace (figure \ref{fig:Scheme-of-replacements})
$\partial{\cal L}/\partial\dot{\mathbf{q}}=\mathbf{p}$ , $\partial{\cal L}/\partial\mathbf{q}=d\mathbf{p}/dt$
, $\partial{\cal L}/\partial t=dH/dt$ and ${\cal L}=\mathbf{p}\dot{\mathbf{q}}-H$.
With these replacements, relation (\ref{eq:lagrangian:development})
is written, to the first order in $\epsilon$, as 
\begin{equation}
\dot{\mathbf{p}}\phi_{\mathbf{q}}+\mathbf{p}\left(\dot{\phi}_{\mathbf{q}}-\dot{\mathbf{q}}\dot{\phi}_{t}\right)-\dot{H}\phi_{t}+\left(\mathbf{p}\dot{\mathbf{q}}-H\right)\dot{\phi}_{t}=0\label{eq:rund:trautman}
\end{equation}
We now only have to notice that the terms in $\mathbf{p}\dot{\mathbf{q}}\dot{\phi}_{t}$
cancel out in (\ref{eq:rund:trautman}), and the remaining terms can
be grouped into 
\begin{equation}
\frac{d}{dt}\left(\mathbf{p}\phi_{\mathbf{q}}-H\phi_{t}\right)=0\label{eq:noether:charge:differential}
\end{equation}
In other words, the quantity between parenthesis in the above expression
is conserved along a trajectory. These cancellations and grouping
to give the final result would have seemed \emph{miraculous} or mysterious,
if we had not, in the previous subsection, given the geometric reason
behind them. 

\section{Applications}

Let us suppose that we are given a certain symmetry and let us study
its consequence, for some simple systems. 

For a purely spatial symmetry where $\phi_{t}=0$, the conservation
law reduces to 
\begin{equation}
\mathbf{p}\boldsymbol{\phi}_{\mathbf{q}}=\text{constant}\label{eq:symmetry:space}
\end{equation}
For a pure translation symmetry where $\boldsymbol{\phi}_{\mathbf{q}}=\mathbf{u}$
(a constant vector), we have simply 
\begin{equation}
\mathbf{p}\mathbf{u}=\sum p_{i}u^{i}=\text{constant}\label{eq:conserve:moment}
\end{equation}
\emph{i.e.} the momentum in the direction of the vector $\mathbf{u}$
is conserved. For a rotation symmetry around the $z$ axis (relations
\ref{eq:rotx},\ref{eq:roty}), the conservation law is 
\begin{equation}
L_{z}=m\left(-y\dot{x}+x\dot{y}\right)=\text{constant}\label{eq:conserv:angular:moment}
\end{equation}
On the other hand, for a pure time translation where $\boldsymbol{\phi}_{\mathbf{q}}=\mathbf{0}$
and $\phi_{t}=a$, we have trivially 
\begin{equation}
H=\text{constant}\label{eq:conserve:H}
\end{equation}

We can use the Noether's theorem in reverse and search for Lagrangians
that are compatible with a given symmetry. Consider for example the
Lagrangians for geodesics in two dimensional Euclidean geometry that
are \emph{strictly} compatible with a rotation symmetry and depend
only on the first derivative : 
\begin{equation}
{\cal L}=f(\dot{y})\label{eq:lagrangian:possible:form}
\end{equation}
where $(x,y)$ are Cartesian coordinates of the plane, $\dot{y}=dy/dx$
($x$ is the independent variable and $y$ the dependent one). For
the rotation symmetry, $x'=x-\epsilon y$, $y'=y+\epsilon x$ and
therefore, $\dot{y}'=\dot{y}+\epsilon(1+\dot{y}^{2})$. Imposing the
strict equality 
\begin{equation}
{\cal L}'dx'={\cal L}dx\label{eq:noether:strict}
\end{equation}
and developing to the first order in $\epsilon$ results in the differential
equation
\begin{equation}
(1+\dot{y}^{2})f'(\dot{y})-\dot{y}f(\dot{y})=0\label{eq:lagrangian:form:ode}
\end{equation}
The solution of the above equation is $f(\dot{y})=C\sqrt{1+\dot{y}^{2}}.$
This form of the Lagrangian  is not surprising, given the fact
that rotations are precisely the transformations that preserve the
usual distance. 

For a Lorentz transformation 
\begin{equation}
x'=\gamma(x-vt)\,\,;\,\,t'=\gamma(t-vx)\label{eq:transform:lorentz}
\end{equation}
where $\gamma=1/\sqrt{1-v^{2}}$, the Infinitesimal generator is $\phi_{x}=-t$,
$\phi_{t}=-x$. Repeating the above computation, one find that for
a free particle, the Lagrangian compatible with Lorentz transform
must be 
\begin{equation}
{\cal L}=C\sqrt{1-\dot{x}^{2}}\label{eq:lorentz:lagrangian}
\end{equation}
The above computations can be generalized without too much difficulty
to Lagrangians of the form 
\begin{equation}
{\cal L}=f(\dot{y},y,t)\label{eq:lorentz:general:lagragian}
\end{equation}
to find the general form of such a function. In particular, Lagrangians
of the form ${\cal L}=a(\dot{y})+\dot{y}b(y,t)+c(y,t)$ will lead
to the general form of Lagrangians for electromagnetic problems compatible
with the requested symmetry. A detailed study of the Noether's theorem
for a charged particle in an electromagnetic field can be found in
\cite{kobe2013}. 

In principle, for a given Lagrangian, we can also look for the transformations
that lead to invariance. This task however is achieved more efficiently
by studying directly the Hamilton-Jacobi equation or by canonical
transformations. Note that by these approaches, invariant quantities
that are not obviously linked to symmetries can be recovered, as for
example in the problem of attraction by two fixed center\cite{waalkens2004}.

\section{Conclusions\protect\label{sec:Conclusions}}

I have discussed in this article the geometrical meaning of the Noether's
theorem and its simple derivation from basics principles. The materials
developed in this short article, which does not contain the usual
mathematical complexities of Rund-Trautman approach found in most
textbooks, can be covered in one lecture and I hope help students
to get a basic understanding of one of the most elegant theorem of
physics. 

\paragraph{Acknowledgment. }

I'm grateful to Cyril Falvo for crtical reading of the manuscript
and fruitful discussions.

\appendix

\section{Miscellaneous results.}

\subsection{Hamiltonian variation as a function of time.\protect\label{subsec:Hamiltonian-variation}}

The Hamiltonian is defined as 
\begin{equation}
H=\mathbf{p}\dot{\mathbf{q}}-{\cal L}\label{eq:hamitonian:def}
\end{equation}
Its variation along a trajectory is given by 
\begin{equation}
\frac{dH}{dt}=\frac{d}{dt}\left(\mathbf{p}\dot{\mathbf{q}}\right)-\frac{d{\cal L}}{dt}\label{eq:dH:dt-1}
\end{equation}
On one hand, we have 
\begin{eqnarray}
\frac{d{\cal L}}{dt} & = & \frac{\partial{\cal L}}{\partial\dot{\mathbf{q}}}\frac{d\dot{\mathbf{q}}}{dt}+\frac{\partial{\cal L}}{\partial\mathbf{q}}\frac{d\mathbf{q}}{dt}+\frac{\partial{\cal L}}{\partial t}\label{eq:dL:dt}\\
 & = & \mathbf{p}\ddot{\mathbf{q}}+\dot{\mathbf{p}}\dot{\mathbf{q}}+\frac{\partial{\cal L}}{\partial t}\label{eq:dL:dt:explicit}
\end{eqnarray}
On the other hand, 
\begin{equation}
\frac{d}{dt}\left(\mathbf{p}\dot{\mathbf{q}}\right)=\mathbf{p}\ddot{\mathbf{q}}+\dot{\mathbf{p}}\dot{\mathbf{q}}\label{eq:dpdotq:dt}
\end{equation}
Therefore, expression (\ref{eq:dH:dt-1}) is reduced to 
\begin{equation}
\frac{dH}{dt}=-\frac{d{\cal L}}{dt}\label{eq:dH:dt:final}
\end{equation}

\subsection{Non-unicity of the Lagrangian and the general Noether's theorem..\protect\label{subsec:Non-unicity}}

Consider two Lagrangians ${\cal L}(\mathbf{q},\dot{\mathbf{q}},t)$
and ${\cal L}^{\dagger}(\mathbf{q},\dot{\mathbf{q}},t)$ related through
\begin{eqnarray}
{\cal L}^{\dagger}(\mathbf{q},\dot{\mathbf{q}},t) & = & {\cal L}(\mathbf{q},\dot{\mathbf{q}},t)+\frac{d}{dt}F(\mathbf{q},t)\label{eq:L:dagger}\\
 & = & {\cal L}(\mathbf{q},\dot{\mathbf{q}},t)+\dot{\mathbf{q}}\frac{\partial F}{\partial\mathbf{q}}+\frac{\partial F}{\partial t}\label{eq:L:dagger2}
\end{eqnarray}
We have 
\begin{equation}
\frac{\partial{\cal L}^{\dagger}}{\partial\dot{\mathbf{q}}}=\frac{\partial{\cal L}}{\partial\dot{\mathbf{q}}}+\frac{\partial F}{\partial\mathbf{q}}\label{eq:lagrangian:with:total}
\end{equation}
and therefore 
\begin{eqnarray}
\frac{d}{dt}\left(\frac{\partial{\cal L}^{\dagger}}{\partial\dot{\mathbf{q}}}\right) & = & \frac{d}{dt}\left(\frac{\partial{\cal L}}{\partial\dot{\mathbf{q}}}\right)+\dot{\mathbf{q}}\frac{\partial^{2}F}{\partial\mathbf{q}^{2}}+\frac{\partial^{2}F}{\partial\mathbf{q}\partial t}\label{eq:L11}\\
\frac{\partial{\cal L}^{\dagger}}{\partial\mathbf{q}} & = & \frac{\partial{\cal L}}{\partial\mathbf{q}}+\dot{\mathbf{q}}\frac{\partial^{2}F}{\partial\mathbf{q}^{2}}+\frac{\partial^{2}F}{\partial\mathbf{q}\partial t}\label{eq:L12}
\end{eqnarray}
As a consequence, 
\begin{equation}
\frac{d}{dt}\left(\frac{\partial{\cal L}^{\dagger}}{\partial\dot{\mathbf{q}}}\right)-\frac{\partial{\cal L}^{\dagger}}{\partial\mathbf{q}}=\frac{d}{dt}\left(\frac{\partial{\cal L}}{\partial\dot{\mathbf{q}}}\right)-\frac{\partial{\cal L}}{\partial\mathbf{q}}\label{eq:total:derivative:consquence}
\end{equation}
Both Lagrangians give rise to the same equations of motion. Note that
for clarity, we have used the vectorial notations instead of expressions
containing explicitly the indexes. Usually, expressions such as (\ref{eq:L:dagger2})
are written as 
\begin{equation}
{\cal L}^{\dagger}(\mathbf{q},\dot{\mathbf{q}},t)={\cal L}(\mathbf{q},\dot{\mathbf{q}},t)+\sum_{j}\dot{q}^{j}\frac{\partial F}{\partial q^{j}}+\frac{\partial F}{\partial t}\label{eq:explicit:notation}
\end{equation}
and for example 
\begin{equation}
\frac{d}{dt}\left(\frac{\partial{\cal L}^{\dagger}}{\partial\dot{q}^{i}}\right)=\frac{d}{dt}\left(\frac{\partial{\cal L}^{\dagger}}{\partial\dot{q}^{i}}\right)+\sum_{j}\frac{\partial^{2}F}{\partial q^{i}\partial q^{j}}\dot{q}^{j}+\frac{\partial^{2}F}{\partial q^{i}\partial t}\label{eq:explicit:notaton:2}
\end{equation}

The non-unicity of the Lagrangian has a consequence in the statement
of the Noether's theorem. In relation (\ref{eq:noether:condition}),
the condition for the theorem to apply was stated as
\begin{equation}
{\cal L}'\frac{dt'}{dt}={\cal L}\label{eq:noether:condition:app}
\end{equation}
to the first order in $\epsilon$. As we saw above, the addition of
a total time derivative does not change the trajectories, therefore
we can now slightly relax the Noether's condition and set the condition
as 
\begin{equation}
{\cal L}'\frac{dt'}{dt}={\cal L}+\epsilon\frac{d}{dt}F(\mathbf{q},t)+o(\epsilon)\label{eq:noether:condition:general}
\end{equation}
where $F$ is an arbitrary function. In this case, the computations
of subsection \ref{subsec:Lagrangian-approach.} can be repeated to
show that the conserved quantity in this case is 
\begin{equation}
I=\mathbf{p}\phi_{\mathbf{q}}-H\phi_{t}-F\label{eq:neother:generalized}
\end{equation}

The geometric derivation of relation follows the computations of
subsection \ref{subsec:H-J-Approach} : Consider again the closed
path ${\cal P}=AA'B'BA$ of figure \ref{fig:int:dS}. As mentioned,
we must have $\oint dS=0$. If we have 
\begin{equation}
\int_{A'}^{B'}dS-\int_{A}^{B}dS=\epsilon\int_{A}^{B}dF\label{eq:int:with:df}
\end{equation}
we also must have 
\begin{equation}
\int_{A}^{A'}dS-\int_{B}^{B'}dS=-\epsilon\int_{A}^{B}dF=-\epsilon\left(F(B)-F(A)\right)\label{eq:int:with:df:2}
\end{equation}
which again leads to the generalized Noether's charge of relation
(\ref{eq:neother:generalized}).

This generalized relation does not introduce anything new. To fix
the idea, consider the two dimensional Lagrangian for movement in
a central field
\begin{equation}
{\cal L}=\frac{m}{2}\left(\dot{x}^{2}+\dot{y}^{2}\right)-U(x^{2}+y^{2})+\frac{d}{dt}F(x,y)\label{eq:example1}
\end{equation}
Without the additional total time derivative, the Lagrangian is invariant
under rotational symmetry $\phi_{x}=-y$, $\phi_{y}=x$ and the conserved
quantity is the angular momentum 
\begin{equation}
L=m(-\dot{x}y+\dot{y}x)\label{eq:angular:momentum}
\end{equation}
With the additional term, the expression of $\mathbf{p}$ changes
(for example, $p_{x}=m\dot{x}+\partial_{x}F$) and computing the Noether's
charge (relation \ref{eq:neother:generalized}) produces exactly the
same angular momentum (relation \ref{eq:angular:momentum}) as the
conserved quantity. 

\subsection{Variation of the function $S(\mathbf{q},t)$ and the Hamilton-Jacobi
equation.\protect\label{subsec:Hamilton-Jacobi-equation}}

\begin{figure}
\begin{centering}
\includegraphics[width=0.75\columnwidth]{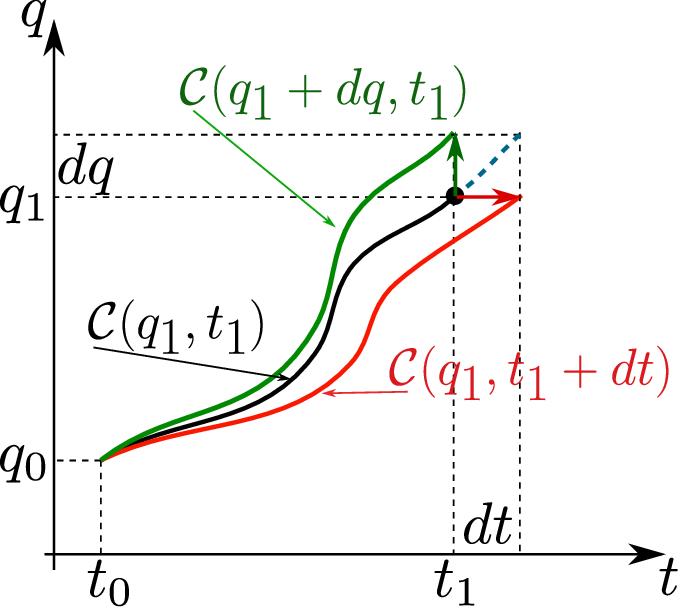}
\par\end{centering}
\caption{Varying the end points of a movement for computation of $\partial_{q}S$
and $\partial_{t}S$.\protect\label{fig:Varying-the-end}}
\end{figure}

Consider an optimal trajectory $\mathbf{q}(t)$ with endpoints \textbf{$(\mathbf{q}_{0},t_{0})$}
and \textbf{$(\mathbf{q}_{1},t_{1})$}. Keeping the initial point
fixed, the function $S(\mathbf{q}_{1},t_{1})$ is the action function.
We are interested in computing the variation of $S()$ when the end
point varies by a small quantity \textbf{$(d\mathbf{q},dt)$ }and
obtain relations (\ref{eq:action:momentum},\ref{eq:action:hamiltonian})
of subsection \ref{subsec:Hamilton-Jacobi-formulation.}.\textbf{
}We begin by keeping the final time fixed at $t_{1}$ but move the
final position by $dq$ (figure \ref{fig:Varying-the-end}). The trajectory
$q(t)$ will vary by $\delta q(t)$ where $\delta q(t_{0})=0$ and
$\delta q(t_{1})=dq$. The variation in $S$ is 
\begin{equation}
\delta S=\int_{t_{0}}^{t_{1}}\left\{ \frac{\partial{\cal L}}{\partial\dot{q}}\delta\dot{q}+\frac{\partial{\cal L}}{\partial q}\delta q\right\} dt\label{eq:deltaS}
\end{equation}
However, the trajectories obey the Euler-Lagrange equation and we
must have 
\[
\frac{\partial{\cal L}}{\partial q}=\frac{d}{dt}\frac{\partial{\cal L}}{\partial\dot{q}}
\]
On the other hand, $\delta\dot{q}=d(\delta q)/dt$. Using these relations,
we can rewrite equation (\ref{eq:deltaS}) as 
\begin{eqnarray*}
\delta S & = & \int_{t_{0}}^{t_{1}}\left\{ \frac{\partial{\cal L}}{\partial\dot{q}}\frac{d\left(\delta q\right)}{dt}+\frac{d}{dt}\left(\frac{\partial{\cal L}}{\partial\dot{q}}\right)\delta q\right\} dt\\
 & = & \int_{t_{0}}^{t_{1}}\frac{d}{dt}\left\{ \frac{\partial{\cal L}}{\partial\dot{q}}\delta q\right\} dt\\
 & = & \left[\frac{\partial{\cal L}}{\partial\dot{q}}\delta q\right]_{t_{0}}^{t_{1}}=\left.\frac{\partial{\cal L}}{\partial\dot{q}}\right|_{t_{1}}dq
\end{eqnarray*}
As we have kept the final time fixed, $\delta S=\left(\partial S/\partial q\right)dq$
and therefore 
\begin{equation}
\frac{\partial S}{\partial q}=\left.\frac{\partial{\cal L}}{\partial\dot{q}}\right|_{t_{1}}=p(t_{1})\label{eq:HJ1}
\end{equation}
If we vary the end point $q_{1}$, the relative variation in $S$
is the\emph{ momentum} $p$ at the end point, as stated in relation
(\ref{eq:action:momentum})

To compute the variation of $S$ as a function of the end point's
time, consider letting the original trajectory to continue along its
optimal path. Then $dS={\cal L}dt.$ On the other hand 
\[
dS={\cal L}dt=\frac{\partial S}{\partial q}dq+\frac{\partial S}{\partial t}dt
\]
Using our previous result (\ref{eq:HJ1}), we have 
\[
{\cal L}dt=pdq+\frac{\partial S}{\partial t}dt=\left(p\dot{q}+\frac{\partial S}{\partial t}\right)dt
\]
and therefore 
\begin{equation}
\frac{\partial S}{\partial t}={\cal L}-p\dot{q}=-H\label{eq:HJ2}
\end{equation}
which is relation (\ref{eq:action:hamiltonian}).

The full derivation of the Hamilton-Jacobi equation can be found in
(\cite{houchmandzadehHamiltonJacobiEquation2020}, we recall here
only its form and use. The function $S(\mathbf{q},t)$ obeys the Hamilton-Jacobi
equation ((HJE) 
\begin{equation}
\frac{\partial S}{\partial t}+H\left(\mathbf{q},\frac{\partial S}{\partial\mathbf{q}},t\right)=0\label{eq:HJE}
\end{equation}
which is a first order partial differential equation in $S$. The
function $H$ is the Hamiltonian, written as a function of $\mathbf{q}$
and the momentum $\mathbf{p}$. Consider for example the 2d harmonic
oscillator (relation \ref{eq:harmonic:osc:hamiltonian}) 
\begin{equation}
H=\frac{m}{2}\left(\dot{x}^{2}+\dot{y}^{2}\right)+\frac{k}{2}\left(x^{2}+y^{2}\right)\label{eq:Hamiltonian:harmonic}
\end{equation}
and $\mathbf{p}=(p_{x},p_{y})=m(\dot{x},\dot{y})$. We rewrite the
Hamiltonian as 
\begin{equation}
H=\frac{1}{2m}\left(p_{x}^{2}+p_{y}^{2}\right)+\frac{k}{2}\left(x^{2}+y^{2}\right)\label{eq:Hamiltonian:harmonic:p}
\end{equation}
The HJ equation for the harmonic oscillator is then
\begin{equation}
\frac{\partial S}{\partial t}+\frac{1}{2m}\left(\left(\frac{\partial S}{\partial x}\right)^{2}+\left(\frac{\partial S}{\partial y}\right)^{2}\right)+\frac{k}{2}\left(x^{2}+y^{2}\right)=0\label{eq:HJE-1}
\end{equation}
Even though HJE could seem complicated, there exist a systematic method
to search for its solution, called \emph{canonical transformations}
(see for example \cite[section 10.4]{goldsteinClassicalMechanics2013}).
When the potential function is separable $V(\mathbf{q})=\sum_{i}V_{i}(q_{i})$,
we can look for a separable solution of the HJE.

\subsection{Fundamental Theorem of calculus.\protect\label{subsec:Fundamental-Theorem}}

The fundamental theorem of calculus of integration of an exact ``differential''
for a multi-variable function over a closed path states that
\begin{equation}
\oint df=0\label{eq:fund:appendix}
\end{equation}
This theorem is known by various names to physics students such as
the Stokes or the gradient theorem which is written in vector calculus
as 
\[
\oint\nabla f.d\mathbf{r}=0
\]
For example, if $f(\mathbf{r})$ is the electric potential, $\nabla f$
is the electric field and its circulation over a closed path is zero. 

The most general approach to its demonstration is through the use
of the language of differential forms\cite{edwardsAdvancedCalculusDifferential1993}
that provides a unified approach to integrals over curves, surfaces
and higher order manifolds. I briefly summarize here the main results
of this field that are related to relation (\ref{eq:fund:appendix}).
For a space referred by coordinates $\mathbf{x=}(x_{1},\ldots x_{n})$,
a 1-form is an object such as 
\begin{equation}
\omega=A_{1}(x_{1},\ldots x_{n})dx_{1}+\ldots A_{n}(x_{1},\ldots x_{n})dx_{n}\label{eq:one-form}
\end{equation}
where $A_{i}(\mathbf{x})$ are functions of $n-$variables. A $2-$form
involves elements such as $dx_{i}dx_{j}$ with the property that 
\[
dx_{i}dx_{j}=-dx_{j}dx_{i}\,\,\,;\,\,\,\,dx_{i}dx_{i}=0
\]
and can be written as 
\begin{equation}
\omega=\sum_{i,j,i\ne j}A_{ij}(\mathbf{x})dx_{i}dx_{j}\label{eq:two-forms}
\end{equation}
The definition can be generalized to $k-$forms. A zero-form is a
scalar function. The \emph{exterior derivative} transforms a $k-$form
into a $(k+1)-$form. For example, 
\[
df=\sum_{i=1}^{n}\frac{\partial f}{\partial x_{i}}dx_{i}
\]
The exterior derivative of a $1-$form such as $\alpha=A(\mathbf{r})dx_{k}$
is 
\[
d\alpha=\sum_{i=1,i\ne k}^{n}\frac{\partial A}{\partial x_{i}}dx_{i}dx_{k}
\]
because, by definition, $dx_{i}dx_{i}=0$. This process can be used
to define the exterior derivation of arbitrary $k-$form. In particular,
for any differential form $\omega$
\begin{equation}
d(d\omega)=0\label{eq:ddw}
\end{equation}
because such a derivation involves only terms of the form 
\[
\frac{\partial^{2}A}{\partial x_{i}\partial x_{j}}-\frac{\partial^{2}A}{\partial x_{j}\partial x_{i}}=0
\]

A differential form that is the derivative of another one is called
an exact form. 

The integration of a differential $k-$form over a domain ${\cal D}$
corresponds to the usual definition of integration of multi-variable
functions. The main theorem of integration of differential forms,
known often as the generalized Stokes theorem states that 
\begin{equation}
\int_{{\cal \partial D}}\omega=\int_{{\cal D}}d\omega\label{eq:form:integration}
\end{equation}
where $\partial D$ is the boundary of ${\cal D}$. Physics students
often encounter specific cases of the above relations in vector calculus
as Stokes and divergence (Ostrogradsky) theorems. For a function of
one variable $f(x)$ integrated over the interval $[a,b]$, the above
relation is the classical 
\[
\int_{a}^{b}df=f(b)-f(a)
\]
Now, consider the exact form $\omega=df$ where $f(\mathbf{x})$ is
a function a $n-$variable and its integration over a closed path
$\partial{\cal D}$ : 
\[
\int_{{\cal \partial D}}df=\int_{{\cal D}}d(df)=0
\]
by the virtue of relation (\ref{eq:ddw}). 


\bibliographystyle{unsrt}

\end{document}